\documentclass{JHEP3}
\title{The confining string beyond the free-string approximation
in the gauge dual of percolation }
\author{Pietro Giudice\\School of Mathematics, Trinity College, Dublin 2, 
Ireland\\
\email{giudice@maths.tcd.ie}}

\author{Ferdinando Gliozzi, Stefano Lottini\\
Dipartimento di Fisica Teorica, Universit\`a di Torino and \\
INFN, Sezione di Torino\\
via P.Giuria 1, I-10125 Torino, Italy\\
\email{gliozzi,lottini@to.infn.it}}
\abstract{We simulate five different systems belonging to the universality 
class of the gauge dual of three-dimensional random percolation to study 
the underlying effective string theory at finite temperature. All 
the data for the finite temperature string tension, when expressed 
by means of adimensional variables, are nicely described  by 
a unique scaling function. We calculate the first few terms of the string 
tension up to order $T^6$ and compare to different theoretical predictions. 
We obtain unambiguous evidence that the coefficients of $T^2$ and 
$T^4$ terms coincide with those of the Nambu-Goto string, as expected, while the $T^6$ term strongly differs and is characteristic of the universality class
of this specific gauge theory.}
\usepackage[dvips]{graphicx}
\usepackage{amsmath}
\usepackage{amsfonts}
\usepackage{mathrsfs}
\usepackage{epsfig}
\usepackage{psfrag}
\newcommand{\Z}{\mathbb{Z}}

\newcommand{\T}{{\cal T}}
\newcommand{\bra}{\langle}
\newcommand{\ket}{\rangle}
\newcommand{\avg}[1]{\langle \hspace{0.2em} #1 \hspace{0.2em} \rangle}
\newcommand{\um}{\frac12}

\newcommand{\eq}{\begin{equation}}
\newcommand{\en}{\end{equation}}
\newcommand{\bea}{\begin{eqnarray}}
\newcommand{\ea}{\end{eqnarray}}

\begin{document}
\section{Introduction}
The possibility of describing the long-distance dynamics of strong interactions
 in the confining phase by an effective string theory is a fascinating, 
many years old conjecture  which dates from before the formulation 
of QCD \cite{first}. In gauge theories it is based on the very 
intuitive assumption 
that the colour flux connecting  a pair of distant quarks is concentrated, 
in the confining phase, inside  a thin flux tube, which then generates the 
linear rising of the confining potential. 
According to the common lore, this thin flux tube should behave, when the 
quarks are pulled very far apart, as a free vibrating string \cite{nambu}.

The string-like nature of the flux tube is particularly evident in the strong 
coupling region, where the vacuum expectation value of large Wilson loops is 
given by a sum over certain lattice surfaces which can be considered as the 
world-sheets of the underlying confining string.
At the roughening point \cite{rough1} this sum diverges and the colour flux tube of whatever three-dimensional or four-dimensional lattice gauge theory 
undergoes a  transition towards a rough phase. It is widely believed that 
such a phase transition of the flux tube belongs to the Kosterlitz-Thouless 
universality class \cite{kt}. Accordingly, the renormalisation group 
equations imply that the effective string action
$S$ describing the dynamics of the flux tube in the whole rough phase 
(the one connected with the continuum limit) flows at large scales towards a 
massless free field theory. Thus, for large enough inter-quark separations 
it is not necessary to know explicitly the specific form of the effective string action $S$, but only its infrared limit
\eq
S[h]=S_{cl}+S_0[h]+\dots,
\label{frees}
\en
where the classical action $S_{cl}$ describes the usual perimeter-area term,
$h$ denotes the two-dimensional bosonic 
fields $h_i(\xi_1,\xi_2)$  with  $i=1,2,\dots, d-2$     
which describe the 
transverse displacements of the string with respect the configuration 
of minimal energy, $\xi_1,\xi_2$ are the coordinates on the world-sheet
and $S_0[h]$ is the Gaussian action
\eq
S_0[h]=\um\int d^2\xi\,\partial_\alpha h_i(\xi_1,\xi_2)\partial^\alpha 
h^i(\xi_1,\xi_2)~~~~ ~~~~(\alpha=1,2;~i=1,2,\dots,d-2)~.
\en
In this IR approximation the effective string is known as the free bosonic 
string. 
The ensuing universal string fluctuation effects \cite{lsw,lu}
were first 
unambiguously observed many years ago in the $\Z_2$ gauge theory in three 
dimensions \cite{Caselle:1992ue,Caselle:1996ii}. 

In order 
to study the first perturbative corrections to the IR limit it has been 
often assumed, for the sake of simplicity, that the effective string action
is  the Nambu-Goto action, i.e. the one proportional to the world-sheet area.
Expanding in the natural dimensionless parameter $1/(\sigma\,A)$, where $\sigma$ is the string tension and $A$ the area of the minimal surface bounded by the 
Wilson loop, one can write
\eq
S[h]=S_{cl}+S_0[h]+\frac1{8\,\sigma A}\int d^2\xi \left[\left(\partial_\alpha 
h_i\partial^\alpha h^i\right)^2- 2\partial_\alpha 
h_i\partial^\beta h^i \partial^\alpha h_j\partial_\beta h^j\right]+
O\left(\frac1{(\sigma A)^2}\right)~. 
\en
Also in this case the first numerical analysis has been performed in a 
3D $\Z_2$ gauge model \cite{Caselle:1994df}. More 
recently, high precision numerical simulations in $SU(N)$ gauge 
theories  confirmed these effects in the static quark 
potential \cite{lw2,maj}. Mismatches between the observed spectrum of the 
low-lying string states  with fixed ends  and the predictions of the free
bosonic  and the Nambu-Goto strings have been repeatedly reported 
 \cite{pemi,kuti,cas1,cas2}. Recent studies demonstrated that these 
mismatches are gradually disappearing at larger distances 
 \cite{HariDass:2006pq,Brandt:2007iw}. 

Closed strings wrapping around a compact dimension \cite{teper} or 
strings at finite temperature \cite{cas3} have also 
been considered.
In this case a remarkable agreement between the observed data and the
 Nambu-Goto predictions has been reported. 
From a theoretical point of view the reasons of this agreement can 
be understood, at least in part, resorting to a systematic expansion
of the most general form of the effective string action $S[h]$ in terms  
 of $h_\alpha(\xi)$ and its derivatives \cite{lw2,lw4}. The outcome of these studies can be conveniently summarised in some general properties of the first few
terms of the low temperature expansion of the string tension
\eq
\sigma(T)=\sigma_0-(d-2)\frac\pi6 T^2+\sum_{n\ge3}s_nT^n~.
\label{sexpa}
\en 
The second term on the right hand side is the low temperature analogue of the 
L\"uscher term of the inter-quark potential \cite{pa}. It is a characteristic 
quantum effect of the  IR free string limit 
(\ref{frees}) and it is expected to be independent of the interaction terms 
of the effective theory. On the side of the gauge theory it is more than 
universal, in the sense that it does not depend on the nature of the 
gauge group. 

As a consequence of a certain open-closed string duality \cite{lw4} it 
was shown that for any number of space-time dimensions $s_3\equiv0$ and that 
in three dimensions $s_4$ is
again a more than universal coefficient which can be evaluated in various 
ways \cite{arvis,df} and coincides with the Nambu-Goto value $s_4^{NG}$  
\eq
s_4=s^{NG}_4=-(d-2)^2\frac{\pi^2}{72\,\sigma_0 }~.
\label{c4}
\en
A different approach to effective string theory \cite{ps} leads to similar 
conclusions \cite{dr,hdm} (i.e. $s_3=0$, $s_2$ and $s_4$ more than universal),
but for all values of $d$.

In spite of the remarkable agreement of the first few terms of the 
Nambu-Goto expansion with the numerical results, theoretical reasons 
indicate that the Nambu-Goto string is a sick theory and  cannot describe 
the effective confining string to all 
orders in $T$: depending on the quantisation method, one finds either the 
breaking of  rotational invariance or  appearance of the 
conformal Liouville mode, in contrast with the assumption that the only physical degrees of freedom of the effective string are the transverse modes. 

Numerical experiments lead to similar conclusions, 
showing that different gauge theories are described, at least at short distance,
by different effective strings \cite{cas1,cas2}, even if, so far, the order 
of the first term deviating from the more than universal behaviour in 
the power expansion of $\sigma(T)$ has not been determined.
In this paper we find the order of such a term 
by evaluating in a particularly simple model, the gauge dual of random 
percolation in three dimensions, the 
coefficients $s_n$ up to $n=6$ order. We find that $s_6$ strongly deviates from 
the value predicted by the Nambu-Goto model. 

We performed five different kinds of 
high-precision numerical experiments by varying the implementation of the 
percolation, the lattice spacing, 
the temporal extent of the lattice and the type of lattice. 
All these variations should keep the system in the same universality class. 
Indeed, as expected, all the 
collected  data agree with the more than universal values of $s_2$ and $s_4$ 
and lead to
\eq
s_5\simeq 0~;~~s_6=\frac{\pi^3}{C\sigma_o^2}~,~~C\simeq300~.
\en 
(See Table \ref{Table:2} for more details). The vanishing 
of $s_5$ suggests that the high temperature expansion is even in $T$, like 
in Nambu-Goto case. Notice however that the value we find for $s_6$ for the gauge dual of percolation is very different from the corresponding coefficient of 
Nambu-Goto string, which is negative: 
$s^{NG}_6=-(d-2)^3\frac{\pi^3}{432\,\sigma_0^2}$.
Preliminary  results have been presented in \cite{Giudice:2007rb,Giudice:2008zk}.

\section{Polyakov loops}

	We focused on the behaviour of the Polyakov-Polyakov correlation 
function at finite temperature in a $(2+1)$-dimensional system.  
The lattice is a $L^2 \times \ell$ slice with periodic boundary conditions, 
with $L$ large enough to 
represent the spatial extent and $\ell = \frac{1}{a\,T}$ the inverse 
temperature. 
We considered a pair of Polyakov loops orthogonal to the spatial direction 
and at a distance of $r$ lattice spacings $a$; the (connected) correlation 
function in this case is denoted by $\avg{P(0)P^*(r)}$.
	
At the free string or leading order (LO) approximation (\ref{frees}) the functional form of this correlator in the effective string picture was calculated 
in different contexts. In lattice gauge theory it was first derived in 
\cite{dfsst}, leading to 
\eq
 \avg{P(0)P^*(r)}_{LO} \propto\frac{ e^{-\sigma\ell r-\mu\ell}}{\eta(\tau)^{d-2}}~,
\en 
where the Dedekind $\eta$ function
is defined as
\eq
	\eta(\tau) \equiv q^{\frac{1}{24}} \prod_{n=1}^\infty (1-q^n)~,~~ 
	\tau =\frac{i \ell}{2 r}~, 
\qquad q\equiv e^{2 \pi i \tau} ~. 
\en
Within this approximation the temperature-dependent string tension, defined as the coefficient of the linear part of the confining potential, i. e.
\eq
\sigma(T)=-\lim_{r\to \infty}\frac1{rT} \log\bra P(0)P^*(r)\ket~,
\label{def}
\en
turns out to be
\eq
\sigma(T)=\sigma-(d-2)\frac\pi6T^2=\sigma_0-(d-2)\frac\pi6T^2+O(T^4)~,
\en
as expected from (\ref{sexpa}); $\sigma_0$ is the zero-temperature 
string tension and $\sigma=\sigma_0 +O(T^4)$. 

At the next to the leading order (NLO) the functional form of the correlator 
has been calculated in \cite{df} 
\eq
\label{pp}
		\bra P(0)P^*(r)\ket_{NLO} = 
\frac{e^{-\mu\ell-\tilde\sigma r\ell}}{\eta(\tau)^{d-2}}\left(1+
\frac{(d-2)\pi^2\ell
[2E_4(\tau)+(d-4)E_2^2(\tau)]}{1152\tilde\sigma r^3}+O(\frac1{r^5})\right)\,,
\en
	where the functions  $E_2$ and $E_4$ (second and fourth Eisenstein 
functions) are defined by:
	\begin{eqnarray}
E_2(\tau)  & \equiv & 1- 24 \sum_{n=1}^\infty \sigma_1(n) q^n \, , \\
E_4(\tau)  & \equiv & 1+240 \sum_{n=1}^\infty \sigma_3(n) q^n \, .
\end{eqnarray}
	The functions $\sigma_i(n)$ here represent the sum of the $i$-th 
powers of all divisors of $n$. 
Using the definition (\ref{def}) one can easily verify that 
the parameter $\tilde\sigma$ is related to $\sigma(T)$ and $\sigma_0$ 
through
\eq
\sigma(T)=\tilde\sigma-\frac\pi6 T^2-\frac{\pi^2}{72\tilde\sigma}T^4=
\sigma_0-\frac\pi6 T^2-\frac{\pi^2}{72\sigma_0}T^4+O(T^5)~.
\label{sigmaT}
\en

\section{The model}
\label{model}
In this work, the model we chose as laboratory to study the effective 
string theory is the three-dimensional random percolation \cite{rpm}. 
It can be seen as the gauge dual of a Q-state Potts model in the limit $Q\to1$.

It is well known that for integer  $Q>1$ one can formulate the gauge Potts 
model either in terms of gauge fields or in the dual version in terms of the 
spin variables. From a computational point of view the latter
is much more convenient in lattice simulations. It is then useful to map 
the needed gauge invariant 
observables (Wilson loops or Polyakov correlators) into the corresponding 
quantities of the dual version and  not to worry about the gauge formulation.
This approach particularly well suits random percolation, as  the direct 
gauge formulation is not (yet) known, but the rules to evaluate the gauge 
invariant observables are unambiguously defined in any configuration of the 
system, as we shall see below.

In the bond percolation model the lattice configurations are generated 
as follows.
Each link of a three-dimensional lattice $\Lambda$ is independently set 
to \emph{on} or \emph{off} according to some fixed probability $p$, which plays the role of a coupling constant. The set of $on$ links, 
or \emph{active links}, forms a graph $G$, whose connected components are 
known as clusters. When $p$
exceeds  a threshold value $p_c$, 
depending on the nature of the lattice,  an infinite, percolating 
cluster forms.      

Similarly, in the site percolation model, another possible formulation of the random percolation, one independently sets \emph{on} or \emph{off} the nodes of 
the lattice with a fixed probability $p$ and generates a graph $G$ by 
putting an active link for each pair of adjacent $on$ nodes.

The key ingredient to extract from the above ensemble of graphs  the 
relevant information on the underlying gauge theory is the definition of the percolation counterpart of the Wilson operator $W_\gamma$ associated with whatever
closed path of the dual lattice. We set $W_\gamma(G)=1$ if there is no path 
of $G$  topologically linked to $\gamma$, otherwise we set $W_\gamma(G)=0$.
In other words, $W_\gamma$ is a projector on the  ensemble of graphs whose image is the subset of graphs not linked to $\gamma$. Therefore its vacuum expectation 
value $\bra W_\gamma\ket$ coincides with the average probability that there is
no path in any cluster linked to $\gamma$.

As in usual gauge theories, evaluating these quantities yields the main physical properties of the model. In this way it has been shown that the percolating phase is confining. The string tension $\sigma$ and the other physical observables have the expected scaling behaviour  dictated by the universality class of 
three-dimensional percolation, therefore such a theory  has a well-defined continuum limit \cite{rpm}. Moreover it has a non-trivial glueball spectrum 
\cite{Lottini:2005ya} and a second-order deconfining transition at finite temperature $T_c$ with a ratio $T_c/\sqrt{\sigma}\simeq1.5$ which turns out to be universal, i.e. it does not depend on the kind of lattice utilised nor on the specific percolation process considered (bond or site percolation).

In the study of the finite size effects described by the effective string 
theory one could use in principle whatever confining gauge theory, owing to the fact that the dominant effects do not depend on the gauge group.
The great advantage of the dual percolation model we study in this paper is 
that its simplicity allows to explore regions that are still inaccessible 
to the other gauge systems from a computational point of view.
\section{Methodology}
In this Section we describe the principal aspects of our method, based on the 
direct measurement of the correlator of two coplanar Polyakov loops at finite
$T<T_c$.  We  begin with the description of the lattice algorithm and proceed to discuss the kind of fits we use to extract the temperature-dependent string tension.
\subsection{Simulations}
We are interested in the universal properties of the effective string  theory 
in the gauge dual of percolation, therefore we studied how  the 
system responds to a variation of the spatial and the temporal sizes of the 
lattices, of the occupancy probability $p$, of the kind of percolation (bond 
or site) and finally of the geometry of the lattice, considering both the simple cubic lattice (SC) and the body-centred cubic lattice (BCC). The set of simulations is listed on Table \ref{Table:1}.
\TABULAR{|ccccc|}{
\hline
Lattice&$p$&$\ell_c=1/aT_c$&temporal sizes $\ell$& spatial sizes \\
\hline 
SC bond& 0.272380&6&$9\div15$&128\\
SC bond& 0.268459&7&$10\div15$&128\\
SC bond& 0.265615&8&$10\div17$&128;194;256;320\\
SC site& 0.3459514&7&$11\div17$&128\\
BCC bond & 0.21113018&3&$4\div10$&128\\
\hline
}
{Relevant parameters of the  simulations.\label{Table:1}}
The values of $p$ are taken from \cite{rpm} and are the occupancy
probabilities corresponding to systems which are at the deconfining temperature
$a T_c=1/\ell_c$ when the (periodic) temporal extension in units of lattice spacing $a$ is the value $\ell_c$ reported in the Table. The simulations were made in the confined phase with a temporal extension in the range 
$\ell_c<\ell\lesssim 3\ell_c$ where the Polyakov-Polyakov correlator is well described by the NLO formula (\ref{pp}). The spatial size was $128^2$ which was in most cases amply sufficient to account for the infinite volume limit. 
Only in two cases, namely $\ell=10$ and $\ell=11$ with $\ell_c=8$, we observed 
a non-negligible dependence on the spatial size. In those cases we performed 
further simulations on  larger  lattices, as indicated on the table, and 
extracted the corrected value of the string tension $\tilde\sigma$ using the 
scaling relation 
\eq
{\tilde{\sigma}}_{1/L}=\tilde\sigma-c\,L^{-1/\nu_2}~,
\en
where $\nu_2=\frac43$ is the thermal exponent of two-dimensional random 
percolation. In both cases the fit to the data was very good. 

       To reach an acceptable statistics, we collected data from $10^5$
configurations for each value of $p$ and $\ell$.
\subsection{Algorithm}
  Due to the particular nature of the random percolation model,
 each configuration can be generated independently from scratch, by
simply filling an empty lattice with links (or sites) that are randomly 
switched on with a probability $p$.

 The tricky part is the measurement of the topological linking of the
 resulting graph $G$ with a pair of Polyakov loops; to this end, we first
 choose a cylindric surface $\Sigma$ bounded by the two loops and look for 
the closed paths of $G$ intersecting it and linked with one of the two 
loops. It is convenient
  to ``clean up'' the graph $G\to G'$, getting rid of dead ends and
bridges between loops, as they cannot belong to 
the mentioned closed paths  \cite{rpm}. 
This is done once for the whole configuration.

        On this ``minimal'' configuration $G'$, then, the surface 
$\Sigma$ is translated in all  possible positions 
and the linking is measured with the technique of reconstructing each
time the clusters in the configuration (by means of the Hoshen-Kopelmann
algorithm) keeping track of the crossings of the loop surface, in order 
to detect nonzero winding numbers.
\subsection{Fits}
The measured Polyakov-Polyakov correlators are compared with the expected 
behaviour (\ref{pp}). Being this an asymptotic expression, valid in the IR 
limit, we  fitted the data to (\ref{pp}) by progressively discarding the 
short distance correlators and taking all the values in the range
$r_{min}\le r\le r_{max}=50 a$, with $r_{min}$   varying from the value $\ell$ 
indicated in the Table \ref{Table:1} to 40 lattice spacings $a$.
The value of the fitted parameter $\tilde\sigma$ as a function of $r_{min}$ 
is plotted in Figure \ref{Figure:1}.
\FIGURE{
\psfrag{sigmatilde}{$~~\tilde\sigma$}
\psfrag{R}{$~r$}
\includegraphics[width=12cm]{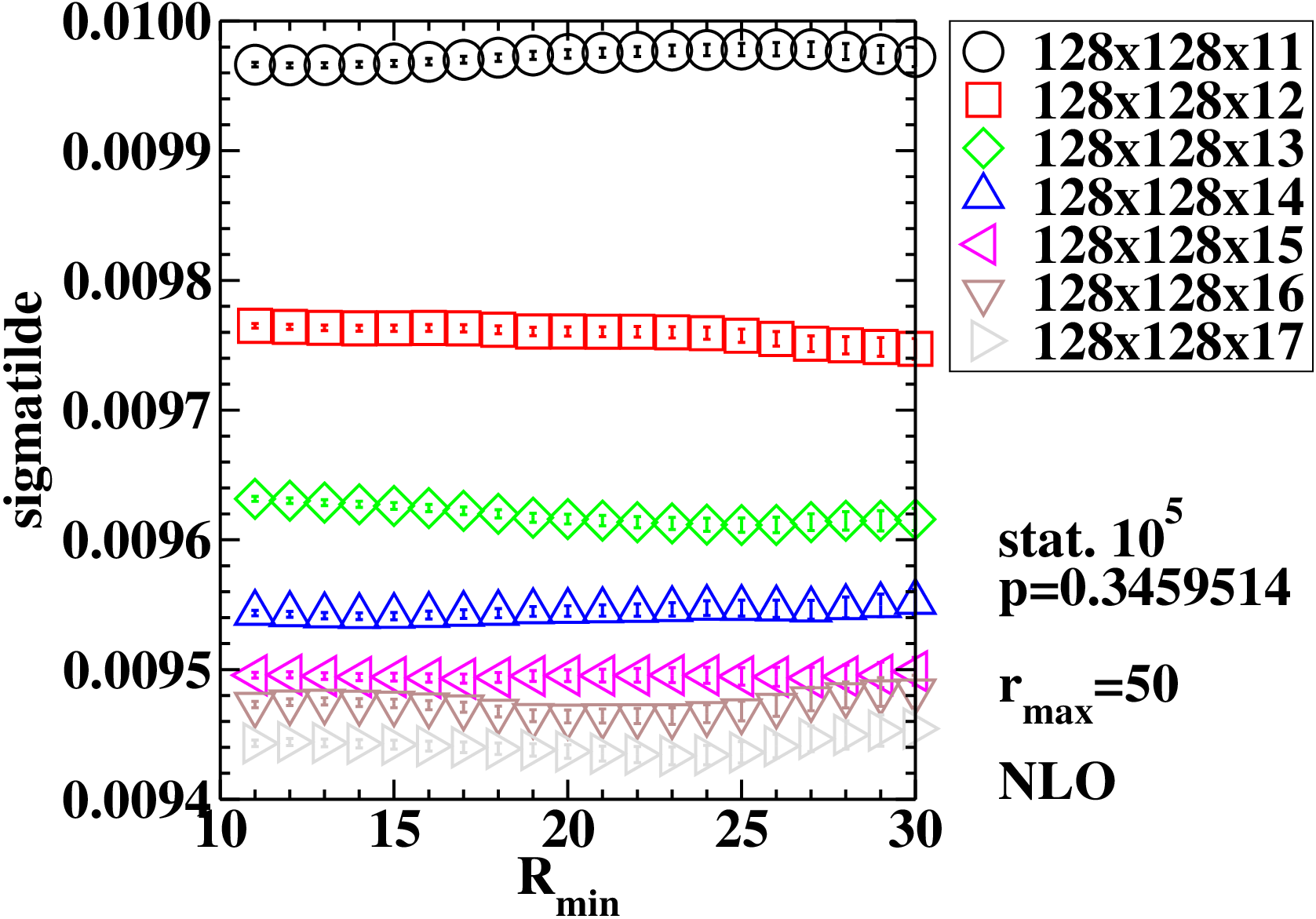}
\caption{The fitted value of $\tilde\sigma$ to (\ref{pp}) as a function of the 
minimal distance $r_{min}$ of the set of Polyakov-Polyakov correlators 
considered in the fit for the site percolation in SC lattice with $\ell_c=7$. 
The different plateaux correspond to different
values of the temporal extension $\ell$. A similar plot for the bond percolation
in the SC lattice  can be found in \cite{Giudice:2007rb} for $\ell_c=6$ and 
in \cite{Giudice:2008zk} for $\ell_c=8$.}
\label{Figure:1}}
The large plateaux in the whole range of the temporal extension $\ell$ 
considered show the stability of the fit which is also supported by a 
$\chi^2/dof$ of the order of 1 or less. In some cases, when $\ell$ is too close
to $\ell_c$, the plateau starts at larger values of $r_{min}$ and correspondingly the $\chi^2$ test is not good. We discarded these data from the further
analysis. In all other cases the Polyakov-Polyakov correlator
in the examined range of $r$ and $l$ is well described by
the asymptotic formula (\ref{pp}). Since the latter is a result of the 
continuum, this agreement can also be interpreted as a check for the absence
of finite lattice spacing effects at the level of our statistical accuracy.

It is important to note that the fitted parameter $\tilde\sigma$ is not yet the 
string tension at zero temperature $\sigma_0$, since (\ref{pp}) is not an
exact formula, but only takes into account the temperature dependence up to the
order $T^4$. On general grounds we expect
\eq
\tilde\sigma=\sigma_0+O(T^5)~.
\label{asys}
\en
 \FIGURE{
\centering
\psfrag{xtemp}{$\times\,\ell$}
\psfrag{temptosix}{$~~T^6$}
\psfrag{sigmatilde}{$~~~\tilde{\sigma}$}
\includegraphics[width=11cm]{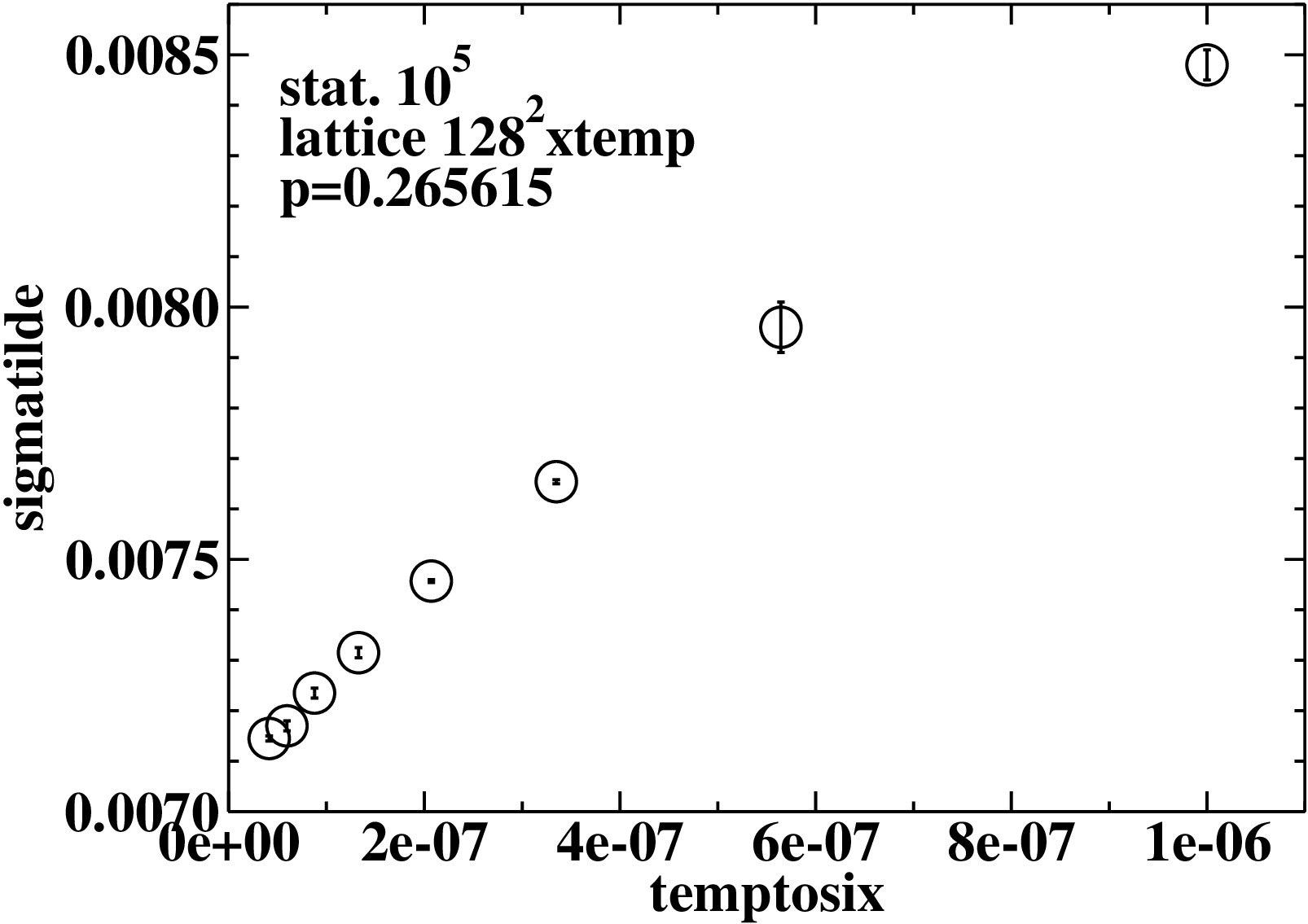}
\caption{Plot of the fitting parameter $\tilde\sigma$ as a function of $T^6$ in 
numerical experiments with bond percolation with $\ell_c=8$. A similar plot for the
case $\ell_c=7$ can be found in \cite{Giudice:2008zk}. }
\label{Figure:2}}
If it turned out that the dependence of the parameter $\tilde\sigma$ 
on $T$ involved lower powers of $T$, i.e. $T^2$ and/or $T^4$, it would mean 
that the first  two thermal corrections in
 (\ref{sigmaT}) were not universal. This question can be settled by studying the dependence on $\ell$ of the mentioned plateaux. In all the cases it turns out that for 
$aT=1/\ell$ low enough the correction is proportional to  $T^6$ (see for instance Figure \ref{Figure:2}). We inserted the fitted parameter $\tilde\sigma$
in (\ref{sigmaT}) in order to reconstruct the quantity $\sigma(T)$ for the whole set of temperatures listed in the fourth column of Table 1.
We then performed, for each line of such a Table, a two-parameter fit to the formula
\eq
\sigma(T)=\sigma_0-\frac\pi6 T^2-\frac{\pi^2}{72\sigma_0}T^4+
\frac{\pi^3}{C\sigma_0^2}T^6+O(T^8)~.
\label{fit}
\en
The fitted parameters $\sigma_0$ and $C$ turn out to be  stable. Their values 
 are reported in Table 2. Another way to analyze the data is to combine 
(\ref{asys}) with the observation that the term $T^5$ is absent and fit 
directly the parameter $\tilde\sigma$ to the formula 
$\tilde\sigma=\sigma_0+\frac{\pi^3}{C\sigma_0^2}T^6$. This way of analysing 
the data differs from the previous one for terms of the order $O(T^8)$, 
thus it can be used for a rough estimate of the systematic errors. 
It turns out that  
the  evaluations of $\sigma_0$ coincide, within the statistical errors, with the values determined in the other way, while the estimates of $C$ are  
about $10\%$ larger than the values reported in Table 2. 

 \section{Results and conclusion}

\FIGURE{
\centering
\psfrag{LL=6 bond SC}{$\ell_c=6$ bond SC}
\psfrag{LL=7 bond SC}{$\ell_c=7$ bond SC}
\psfrag{LL=8 bond SC}{$\ell_c=8$ bond SC}
\psfrag{LL=7 site SC}{$\ell_c=7$ site SC}
\psfrag{LL=3 bond BCC}{$\ell_c=3$ bond BCC}
\psfrag{Nambu-Goto}{Nambu-Goto}
\psfrag{ascissa}{$T/\sqrt{\sigma_0}$}
\psfrag{ordinata}{$\sigma(T)/(T\sqrt{\sigma_0})$}
\includegraphics[width=14cm]{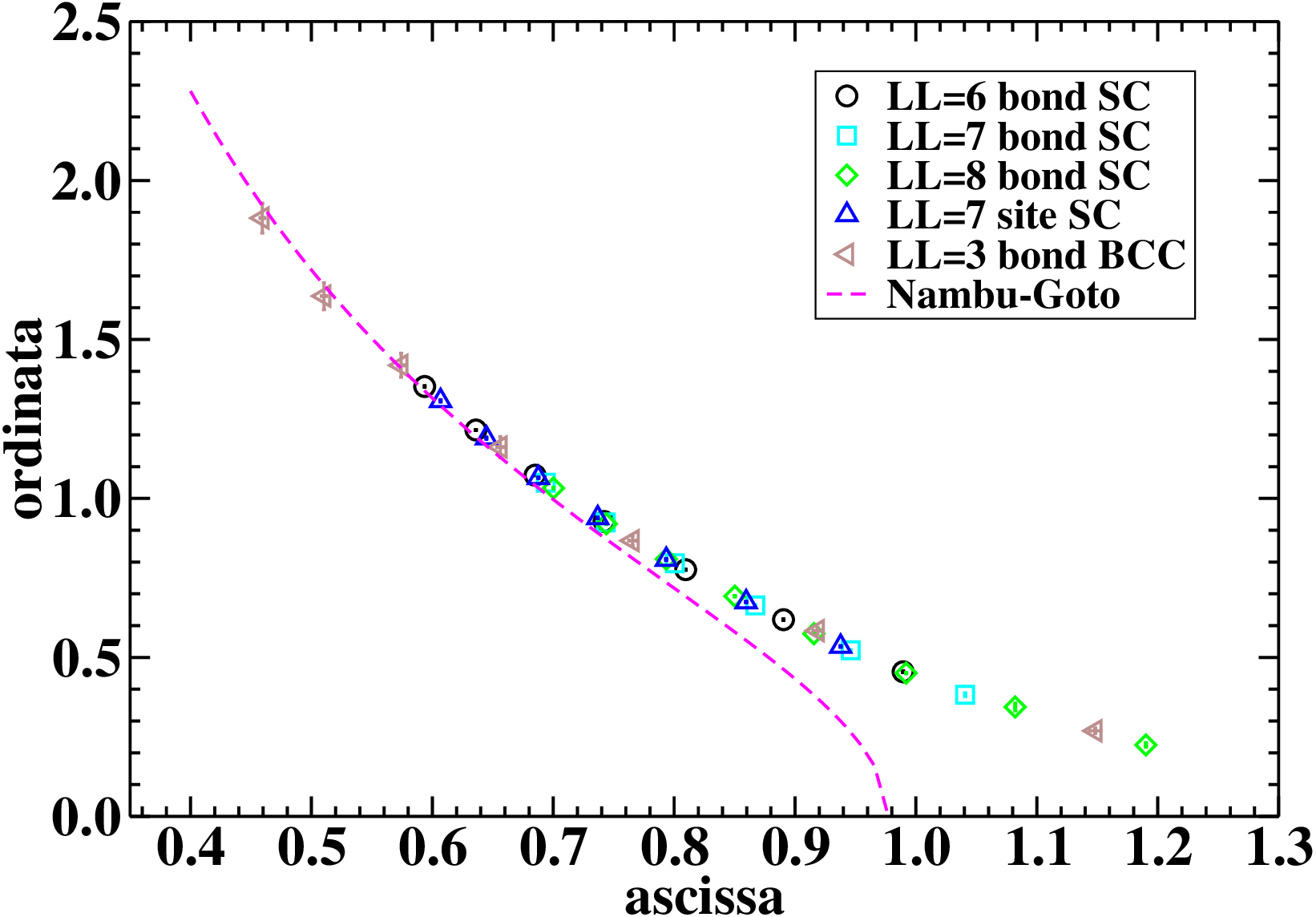}
\caption{Plot of the scaling variable $\sigma(T)/(T\sqrt{\sigma_0})$ as a 
function of $T/\sqrt{\sigma_0}$. The dashed line is the result of the 
Nambu-Goto string.}
\label{Figure:3}}
In this paper we combined Monte Carlo simulations with different finite-size 
scaling techniques applied to various percolating systems. 
The outcome of the extensive numerical experiments on the gauge dual of random percolation and the analysis 
described in the previous Section is a precise determination of the string 
tension as a function of the temperature  in a wide range of   $T$.

If we plot the adimensional ratio $\sigma(T)/(T\sqrt{\sigma_0})$ versus
the adimensional temperature $T/\sqrt{\sigma_0}$ it turns out that all the 
data neatly lie on a unique universal curve as Figure \ref{Figure:3} shows. 
This scaling behaviour indicates that the most relevant sources of systematic errors, including the approach to the infinite volume and the continuum limits, 
have been taken into account. The plotted quantity is expected to vanish at $T_c$ with the power law $\sim (T_c-T)^{\nu_2}$, where $\nu_2=\frac43$ is the thermal exponent of 2D percolation. Unfortunately our data are not sufficiently close to $T_c$ in order to check accurately this behaviour. A similar scaling function has been 
determined for the 3D $SU(2)$ gauge model \cite{mteper}.  

From each set of
the  numerical simulations described in each row of Table \ref{Table:1} we can 
extract three physical quantities. The first one is the coefficient $C$ of 
Eq.~(\ref{fit}) which determines the $T^6$ correction to the string tension. 
The  five values of $C$ generated by as many different systems (see Table 
\ref{Table:2}) 
remarkably coincide up to the statistical errors. Each set of 
simulations yields also a precise determination of $a^2\sigma_0$ which, 
combined with the precise value of the deconfinement temperature in the same 
lattice units, yields the 
adimensional ratio $T_c/\sqrt{\sigma_0}$. This quantity is expected to be constant  in the continuum limit.
        
\TABULAR{|cccccc|}{
\hline
Lattice&$\ell_c=1/aT_c$&$C$&$a^2\sigma_0$&$\chi^2/dof$&
$T_c/\sqrt{\sigma_0}$\\
\hline 
SC bond&6&291(7)&0.012612(6)&0.15&1.4841(4)\\
SC bond&7&281(5)&0.009234(5)&1.2&1.4866(5)\\
SC bond&8&297(5)&0.007059(5)&0.4&1.4878(5)\\
SC site&7&307(9)&0.009399(8)&0.2&1.4735(6)\\
BCC bond&3&295(14)&0.0474(4)&0.8&1.531(7)\\
\hline
}
{The parameter $C$ and $a^2\sigma_0$ in the fit (\ref{fit}) and $\chi^2/dof$ 
which are obtained for the corresponding numerical experiments listed in 
Table \ref{Table:1}. The last column is the universal ratio
$T_c/\sqrt{\sigma_0}$ as obtained by combining the second and the fourth
columns. \label{Table:2}}
 
In order to extrapolate to this limit, one has to take into account the correction to scaling terms. The string tension in the gauge dual of percolation 
is expected to obey the scaling behaviour \cite{rpm}
\eq
a^2\sigma(p)=S(p-p_c)^{2\nu}\left(\frac1{1+B(p-p_c)^{\omega\nu}}\right)~,
\en
where $p_c$ is the critical threshold and $\nu$ and $\omega$ are the thermal and correction-to-scaling exponents of 3D percolation (see \cite{balle} for an accurate numerical estimate of these exponents). Similarly, the deconfining temperature $T_c$ is expected to scale as
\eq
a\,T_c=\T(p-p_c)^{\nu}\left(\frac1{1+C(p-p_c)^{\omega\nu}}\right)~.
\en
When applied to the case of bond percolation in the SC lattice they yield
$S=9.29(2)$ and $\T=4.562(1)$, thus the extrapolated  continuum limit 
of $T_c/\sqrt{\sigma_0}$ is estimated to be $\T/\sqrt{S}=1.497(2)$.

In conclusion, in this paper we extracted from various three-dimensional 
percolating systems some general information on 
the effective string theory describing the infrared properties of the confining 
phase of the gauge dual of percolation at finite temperature. We  
numerically  evaluated the universal scaling function describing the string 
tension as a function of the temperature. We obtained clear evidence that the 
first two non-vanishing coefficients of the expansion of $\sigma$ in powers of 
$T$ coincide with those of the Nambu-Goto string, while the third one strongly 
differs. Nonetheless this term does not depend on the UV cut-off nor on the specific  percolation model, but is characteristic of the universality class of (the gauge 
dual of) the three-dimensional random percolation.

\acknowledgments
FG thanks the Galileo Galilei Institute for Theoretical Physics for hospitality and the INFN for partial support during the Workshop on Non-Perturbative Methods in Strongly Coupled Gauge Theories. He also thanks various participants throughout the workshop for  fruitful discussions, in particular M. d'Elia and 
M. Teper. We would also like to thank M. Caselle for many useful discussions and comments.

\end{document}